\documentclass[12pt]{article}
\usepackage{graphicx}

\def\madison{Wisconsin IceCube Particle Astrophysics Center and Department of Physics\\
University of Wisconsin--Madison, Madison, WI}

\def\Title#1{\begin{center} {\Large #1 } \end{center}}
\def\Author#1{\begin{center}{ \sc #1} \end{center}}
\def\Address#1{\begin{center}{ \it #1} \end{center}}

\newenvironment{Presented}{\begin{quotation} \begin{center} 
             PRESENTED AT\end{center}\bigskip 
      \begin{center}\begin{large}}{\end{large}\end{center} \end{quotation}}
\def\Acknowledgements{\bigskip  \bigskip \begin{center} \begin{large}
             \bf ACKNOWLEDGEMENTS \end{large}\end{center}}
\def\beq{\begin{equation}}
\def\eeq#1{\label{#1}\end{equation}}
\def\eeqn{\end{equation}}

\def\beqa{\begin{eqnarray}}
\def\eeqa#1{\label{#1}\end{eqnarray}}
\def\eeqan{\end{eqnarray}}

\let\bar=\overbar

\def\Dslash{\not{\hbox{\kern-4pt $D$}}}
\def\dslash{\not{\hbox{\kern-2pt $\del$}}}

\def\msb{{\bar{\ssstyle M \kern -1pt S}}}

\begin{document}
\begin{titlepage}
%%\pubblock

\vfill
\Title{CosPA2013: Outlook}
\vfill
\Author{ Francis Halzen}
\Address{\madison}
\vfill
%\begin{Abstract}
%??? Abstract goes here...
%\end{Abstract}
\vfill
\begin{Presented}
The 10th International Symposium
on Cosmology and Particle Astrophysics (CosPA2013)\\
Honolulu, Hawai'i,  November 12--15, 2013
\end{Presented}
\vfill
\end{titlepage}
\def\thefootnote{\fnsymbol{footnote}}
\setcounter{footnote}{0}

This conference celebrated the great breakthroughs achieved in particle and astroparticle physics and, especially, in cosmology in the last decade. An increase in the sensitivity of our instruments, by one or more orders of magnitude, made this possible: accelerators, microwave background detectors, satellite-borne and ground-based gamma ray detectors, gravitational wave interferometers, cosmic ray arrays, underground dark matter and neutrino detectors, neutrino telescopes, and more. Also, the collection and analysis of data sets from astronomical telescopes of a size and complexity previously unimaginable helped transform cosmology to the precision science it is today~\cite{cunha}. We have to keep this up on all fronts, and where prohibitively expensive, we should gamble on ingenuity and risky experiments. What we must avoid is endless debate, which leads to conservatism. We know that the future is promising; unlike previous generations, we cannot live under the illusion that all physics has already been discovered. We have dark matter, dark energy, and neutrino mass to remind us of this, and the discovery of the Higgs particle puts the necessity of new physics at the TeV scale in sharper focus.

With dark matter \cite{bindi} and dark energy \cite{senatore}, astronomers have raised physics problems that appear as daunting now as the problem of the sun's lifetime seemed over one century ago. At the time, evolution and geology required a sun that was older than several tens of millions of years. Neither chemical nor gravitational energy could accommodate this long lifetime, and neither chemistry nor astronomy resolved the puzzle; Becquerel did with the discovery of a new source of energy (in his desk drawer!), radioactivity. History may repeat itself with heavenly problems resolved by Earth-bound experiments, or vice versa, of course. In this context, we should not forget that cosmic accelerators deliver the highest energy protons, photons, and neutrinos that we observe. With the operation of a generation of totally novel detectors \cite{minakata}, cosmic beams may, as was once the case in the pioneering days of cosmic ray physics, yield particle physics results.

I will review two recent breakthroughs in the field that create new opportunities for discovery: the observation of the Higgs boson and the first detection of neutrinos originating beyond the sun. 

\section{The Higgs exists}

To find the Higgs particle, the worldwide particle physics community coordinated a scientific effort, centered on the LHC, on a scale similar to that of the largest space missions. The technical delivery has been striking, and the scientific harvest has been impressive, culminating in the discovery of the Higgs. The impact is immediate: the particle ``seen" through a decade of precision measurements testing electroweak theory actually exists. We now know that the Higgs is not a convenient parametrization for new physics that is not part of electroweak theory; the particle that caps the Standard Model is real. The discovery has a sinister side, though: it brings our ignorance of the origin of symmetry breaking to center stage.

The problem is as old as the weak interaction: why are the weak interactions weak? Though unified with electromagnetism, weak interactions are not apparent in daily life while electromagnetism is. Already in 1934, Fermi provided a phenomenological structure to frame the question, which also laid the groundwork for building the Standard Model. Fermi's theory prescribed a quantitative relation between the fine-structure constant and the weak coupling: $G \sim \alpha /  M_W^2$. Fermi effectively adjusted $M_W$ to accommodate the observed strength and range of nuclear radioactive decays; one can, for instance, readily obtain a value of $M_W$ of 40 GeV from the observed decay rate of the muon for which the proportionality factor is $\pi \over \sqrt2$ in Fermi's original theory. The answer is off by a factor of 2 because the discoveries of parity violation and neutral currents were in the future and these introduce an additional factor $1-  M_W^2 / M_Z^2$:
\begin{equation}
G_\mu =\left[\pi \alpha \over \sqrt2 M_W^2\right]\, \left[1 \over {1 - M_W^2 / M_Z^2}\right]\, ( 1 + \Delta r)\,.
\label{Fermi}
\end{equation}
Fermi could certainly not have anticipated that we now have a renormalizable gauge theory that allows us to calculate the radiative corrections $\Delta r$ to his formula. Besides regular higher order diagrams, loops associated with the top quark and the Higgs boson contribute; their contribution has been measured at LEP.

I once heard one of my favorite physicists refer to the Higgs as the ``ugly" particle; this is nowadays politically incorrect but more true than ever. Indeed, scalar particles are in some sense ``unnatural."  If one calculates the radiative corrections to the mass $m$ appearing in the Higgs potential ${1\over 2} m^2 H^\dagger H + {1 \over 4} \lambda(H^\dagger H)^2$, the same gauge theory that withstood the onslaught of precision experiments at LEP/SLC and the Tevatron reveals loop contributions that grow quadratically:
\begin{eqnarray}
%%\delta m^2 & =& {3\over 16\pi^2 v^2}\left(2m_W^2 + m_Z^2 + m_H^2 - 4 m_t^2\right) \nonumber\\ &&\quad  {}\times \Lambda^2\ , 
\delta m^2 = {3\over 16\pi^2 v^2}\left(2m_W^2 + m_Z^2 + m_H^2 - 4 m_t^2\right) \times \Lambda^2\, 
\label{quadrdiv}
\end{eqnarray}
where $m_H^2=2\lambda v^2$, $\lambda$ is the quartic Higgs coupling, $v=246$ GeV, and $\Lambda$ a cutoff. The growing loop corrections represent a dangerous contribution to the Higgs vacuum expectation value, which eventually destabilizes the beautiful predictions of the theory. From the optimistic point of view, the Standard Model works amazingly well by fixing $\Lambda$ at the electroweak scale, and we can interpret this as an indication of the existence of new physics beyond the Standard Model that will come to the rescue; following Weinberg
\begin{eqnarray}
{\cal L}(M_W) &=& {1\over 2} m^2 H^\dagger H + {1 \over 4} \lambda(H^\dagger H)^2 + {\cal L}^{\rm gauge}_{\rm SM} 
 \nonumber\\
&+&  {\cal L}^{\rm Yukawa}_{\rm SM} + {1\over\Lambda}{\cal L}^5 +  {1\over\Lambda}{\cal  L}^6+ ... .
\label{weinberg}
\end{eqnarray} 
The operators of higher dimension parametrize physics beyond the Standard Model, and just as Fermi anticipated electroweak particle physics at a scale $m_W$ in 1934, the electroweak gauge theory requires new physics to tame the divergences associated with the Higgs potential.  By the most conservative estimates this new physics is within our reach. This can be seen as follows: avoiding fine-tuning beyond the 10\% level, or
\begin{equation}
\label{quadrft}
\left|{\delta m^2 \over m^2}\right|=
\left|{\delta v^2 \over v^2}\right|
\leq 10
\end{equation}
%
%requires $\Lambda \lesssim 2{\sim}3$\,TeV to be revealed by the LHC.
requires a value of $\Lambda \sim{2-3} $\,TeV to be revealed by the LHC.

Over the years, dark clouds have built up and covered this sunny horizon because some electroweak precision measurements match the Standard Model predictions with too high a precision, some pushing $\Lambda$ towards 10\,TeV.  Incidentally, the measured value of the Higgs mass now excludes the possibility that the factor $(2  M_W^2+M_Z^2+ M_h^2-4M_t^2)$ multiplying the unruly quadratic correction vanishes; the Veltman condition is not realized in Nature. 

The conclusion has been reinforced that new physics must appear at $2{\sim}3$ TeV, even though higher scales can be rationalized when accommodating selected experiments. Supersymmetry is a textbook possibility. Even though it elegantly controls the quadratic divergence by the cancellation of boson and fermion contributions, it is also fine-tuned at a scale of $2{\sim}3$\,TeV. There has been an explosion of creativity to resolve the challenge in other ways; the good news is that all involve new physics in the form of scalars, new gauge bosons, or non-standard interactions. Alternatively, it is possible that we may be guessing the future while holding too small a deck of cards and LHC will open a new world that we did not anticipate. Particle physics would return to its early traditions where experiment leads theory, as it should be, and where innovative techniques introduce new accelerators and detection methods that allow us to observe with an open mind and without a plan, leading us to unexpected discoveries. We certainly should not despair. The LHC has not exhausted its discovery potential with its future holding an increase in energy by a factor of two and in luminosity by more than an order of magnitude.

Measuring the properties of the Higgs itself may be a gateway to new physics, although one should realize that, with the present precision, the LHC Higgs measurements are not competitive with precision electroweak tests already performed elsewhere. This can be fixed by building a linear collider.

\section{Neutrinos}

A decade ago, a string of fundamental experimental measurements established the three-flavor framework of oscillating massive neutrinos. At the time, it could be elegantly summarized by three neutrino states, with $e$, $\mu$ and $\tau$ flavors produced in weak processes propagating as mixed states $\nu_1$, $\nu_2$ and~$\nu_3$:
\begin{eqnarray}
\nu_1 &=& -\cos\theta\nu_e + \sin\theta \left(\nu_\mu-\nu_\tau\over\sqrt2\right) \,, \nonumber \\
\nu_2 &=&  \sin\theta\nu_e +  \cos\theta\left(\nu_\mu-\nu_\tau\over\sqrt2\right) \,, \nonumber \\
\nu_3  &=& \left( \nu_\mu + \nu_\tau \over \sqrt2 \right)\,. \label{mixing}
\end{eqnarray}
Here $\theta$ is the solar mixing angle. The discovery of neutrino oscillations in the solar and atmospheric beams has been confirmed by supporting evidence from reactor and accelerator experiments. A new generation of experiments has falsified the assumption that there is no mixing that couples the solar and atmospheric oscillations by measuring a rather large value of the $\theta_{13}$ angle with impressive precision.

Next-generation neutrino experiments will be a lot more challenging. Construction of the KATRIN spectrometer measuring neutrino mass to 0.02 eV by studying the kinematics of tritium decay is in progress, and a wealth of ideas on double beta decay and novel long-baseline experiments are approaching reality. These experiments will have to answer the great ``known-unknowns" of neutrino physics: their absolute mass and hierarchy, the CP-violating phase, and whether neutrinos are really Majorana particles.

The observation of neutrinoless double beta decay would be especially rewarding. Its observation would confirm the theoretical bias that neutrinos are Majorana particles, yield critical information on the absolute mass scale, and, possibly, resolve the hierarchy problem. In the meantime, we will keep wondering whether small neutrino masses are our first glimpse at grand unified theories via the see-saw mechanism, or if they represent a new physics scale tantalizingly connected to lepton conservation and, possibly, the cosmological constant, reborn in the $\Lambda$CDM model of cosmology. 

The cosmological constant represents a thorny issue for the Standard Model. New physics is required to control the Standard Model calculation of the vacuum energy, also known as the cosmological constant, which diverges as
\begin{equation}
\int^\Lambda  {1\over2}\hbar\omega = \int^\Lambda {1\over2}\hbar\sqrt{k^2+ m^2\,}\,  d^2k \ \sim \ \Lambda^4 \,.
\end{equation}
The cutoff energy required to accommodate its ``observed'' value is $\Lambda = 10^{-3}$\,eV, on the order of the neutrino mass.

Information on neutrino mass has emerged from an unexpected direction: cosmology. The structure of the Universe is dictated by the physics of cold dark matter, and the galaxies we see today are the remnants of relatively small overdensities in the nearly uniform distribution of matter in the very early Universe. Overdensity means overpressure that drives an acoustic wave into the other components making up the Universe: the hot gas of nuclei and photons and the neutrinos. These acoustic waves are seen today in the temperature fluctuations of the microwave background as well as in the distribution of galaxies on the sky.  With a contribution to the Universe's matter balance similar to that of light, neutrinos play a secondary role. The role is however identifiable---neutrinos, because of their large mean-free paths, prevent the smaller structures in the cold dark matter from fully developing, and this is visible in the observed distribution of galaxies. Simulations of structure formation with varying amounts of matter in the neutrino component, i.e., varying neutrino mass, can be matched to a variety of observations, including measurements of galaxy-galaxy correlations and temperature fluctuations on the surface of last scattering. Next-generation experiments will probe neutrino mass beyond the reach of KATRIN. 

With time, neutrino physics increasingly appears as yet another area of particle physics that has moved from discovery to precision science. The activity focusing on the search for sterile neutrinos, whether rationally justified by ``anomalies" in reactor, oscillation and cosmological measurements or not, should remind us however that neutrino experiments are still discovery experiments as well.

\section{Discovery of cosmic neutrinos}

\begin{figure}[htpb]
\begin{center}
%%\vspace{-25pt}
\includegraphics[width=0.95\textwidth,trim=0px 70px 0px 10px,clip=true]{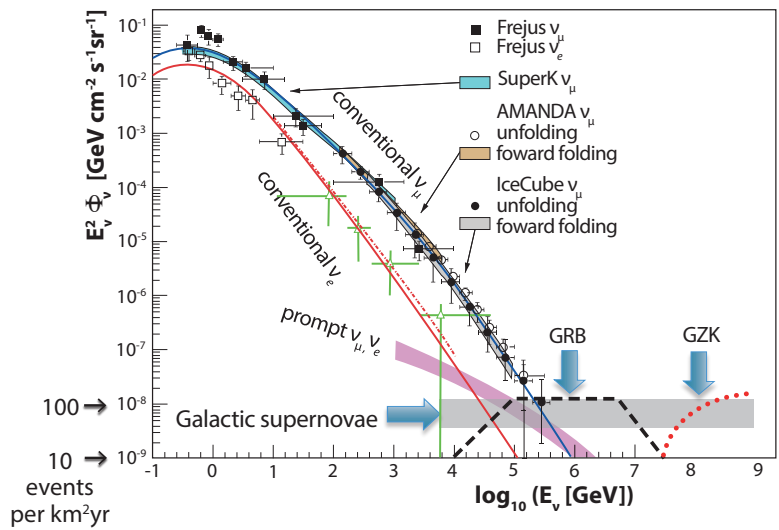}
\caption{Anticipated cosmic-neutrino fluxes produced by supernova remnants and GRBs exceed the atmospheric neutrino flux in IceCube above 100\, TeV. Also shown is a sample calculation of the cosmogenic neutrino flux. The atmospheric electron-neutrino spectrum is shown with green open triangles. The conventional $\nu_e$ (red line) and $\nu_\mu$ (blue line) from Honda, $\nu_e$ (red dotted line) from Bartol and charm-induced neutrinos (magenta band) are shown. Previous measurements from Super-K, Frejus, AMANDA and IceCube are also shown.}
\label{discovery}
\end{center}
\end{figure}

Exactly fifty years ago pioneering experiments in deep underground mines in India and South Africa discovered atmospheric neutrinos. The atmospheric neutrino beam was later exploited by a new generation of underground detectors to demonstrate that neutrinos have a tiny mass representing the first chink in the armor of the Standard Model. In contrast, the detectors' search for cosmic neutrinos reaching us from sources beyond the sun came up empty, establishing an upper limit on their flux; assuming an $E^{-2}$ energy dependence:
\begin{equation}
E_\nu^2 \frac{dN}{dE_\nu} \leq 5\times10^{-6}\,\rm GeV\,cm^{-2}\,s^{-1}\,sr^{-1}
\label{eq:frejus}
\end{equation}
Enter AMANDA and its successor IceCube, which transformed a cubic kilometer of natural Antarctic ice into a neutrino detector. Operating for almost one decade, the AMANDA detector improved this limit by two orders of magnitude. With data taken during its construction, IceCube's sensitivity rapidly approached the neutrino flux level predicted for theorized sources of cosmic rays, such as supernova remnants, gamma-ray bursts and active galactic nuclei; see Fig.~\ref{discovery}. With its completion, IceCube also positioned itself for observing the much anticipated cosmogenic neutrinos, with some estimates predicting as many as two events per year. Cosmogenic neutrinos are produced in the interactions of cosmic rays with microwave photons.

\begin{figure}[htpb]
\begin{center}
%%\vspace{25pt}
\includegraphics[width=0.5\textwidth]{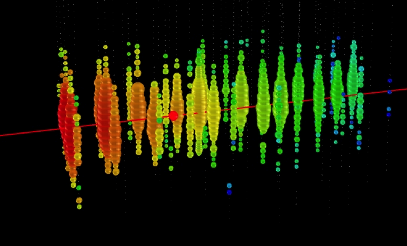}\,\,\,\,\includegraphics[width=0.4\textwidth]{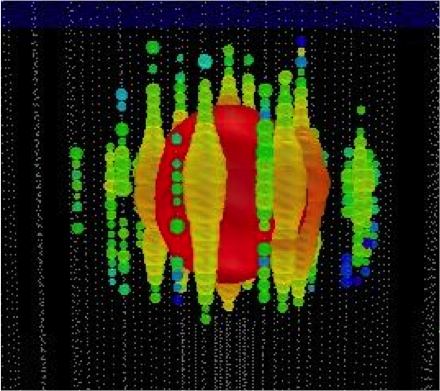}
\caption{Left: a neutrino-induced muon from below the horizon crossing the detector.
Right: a cascade event starting in the detector.
 Color of the dots indicates arrival time, from red (early) to purple (late) following the rainbow. Size of the dots indicates the number of photons detected.}
\label{ApollonBert}
\end{center}
\end{figure}

Cosmogenic neutrinos were the target of a dedicated search using IceCube data collected between May 2010 and May 2012. Two events were found. However, their energies, rather than $\sim$EeV, as expected for cosmogenic neutrinos, were just above one PeV. These events were particle showers initiated by neutrinos interacting inside the instrumented detector volume. Their light pool of roughly one hundred thousand photoelectrons extended over more than 500 meters, as illustrated in the right panel of Fig.~\ref{ApollonBert}. With no evidence of a muon track, they were both most likely initiated by electron or tau neutrinos.

Before this serendipitous discovery, neutrino searches had almost exclusively targeted muon neutrinos that interact primarily outside the detector to produce kilometer-long muon tracks passing through the instrumented volume.  This approach maximizes the event rate by enlarging the target volume, but it is necessary to use the Earth as a filter to remove the huge background flux of muons produced by cosmic ray interactions in the atmosphere. This limits the neutrino view to a single flavor and half the sky. Inspired by the observation of the two PeV events, a filter was designed to identify high-energy neutrinos interacting inside the detector. It divides the instrumented volume of ice into an outer veto shield and a 420 megaton inner fiducial volume. The separation between veto and signal regions was optimized to reduce the background of atmospheric muons and neutrinos to a handful of events per year while keeping 98\% of the signal. The background of atmospheric muon punch-through was determined experimentally by measuring the rate at which muons tagged in the veto region passed an inner veto region of similar 
size.  The great advantage of specializing to neutrinos interacting inside the instrumented volume of ice is that the detector functions as a total absorption calorimeter measuring deposited energy with 10--15\% resolution. Also, neutrinos from all directions in the sky can be identified, including both muon tracks produced in $\nu_\mu$ charged-current interactions and secondary showers produced by neutrinos of all flavors.

Analyzing data covering the same time period as the cosmogenic neutrino search, 28 events were identified with in-detector deposited energies between 30 and 1140 TeV.  Predominantly originating in the Southern Hemisphere, none show evidence for additional muon tracks, thus reinforcing the case that they are extraterrestrial. If atmospheric in origin, the neutrinos should indeed be accompanied by muons produced in the air shower in which they originate. For example, at 1 PeV, less than 0.1\% of atmospheric showers contain no muons with energy above 500 GeV, approximately that which is needed to reach the detector in the deep ice when traveling vertically.

\begin{figure}[htpb]
\begin{center}
%%\vspace{-15pt}
%\includegraphics[width=1.0\linewidth,trim=0px 0px 0px -30px,clip=true]{hesedata.pdf}
\includegraphics[width=0.9\linewidth,trim=0px 0px 0px -30px,clip=true]{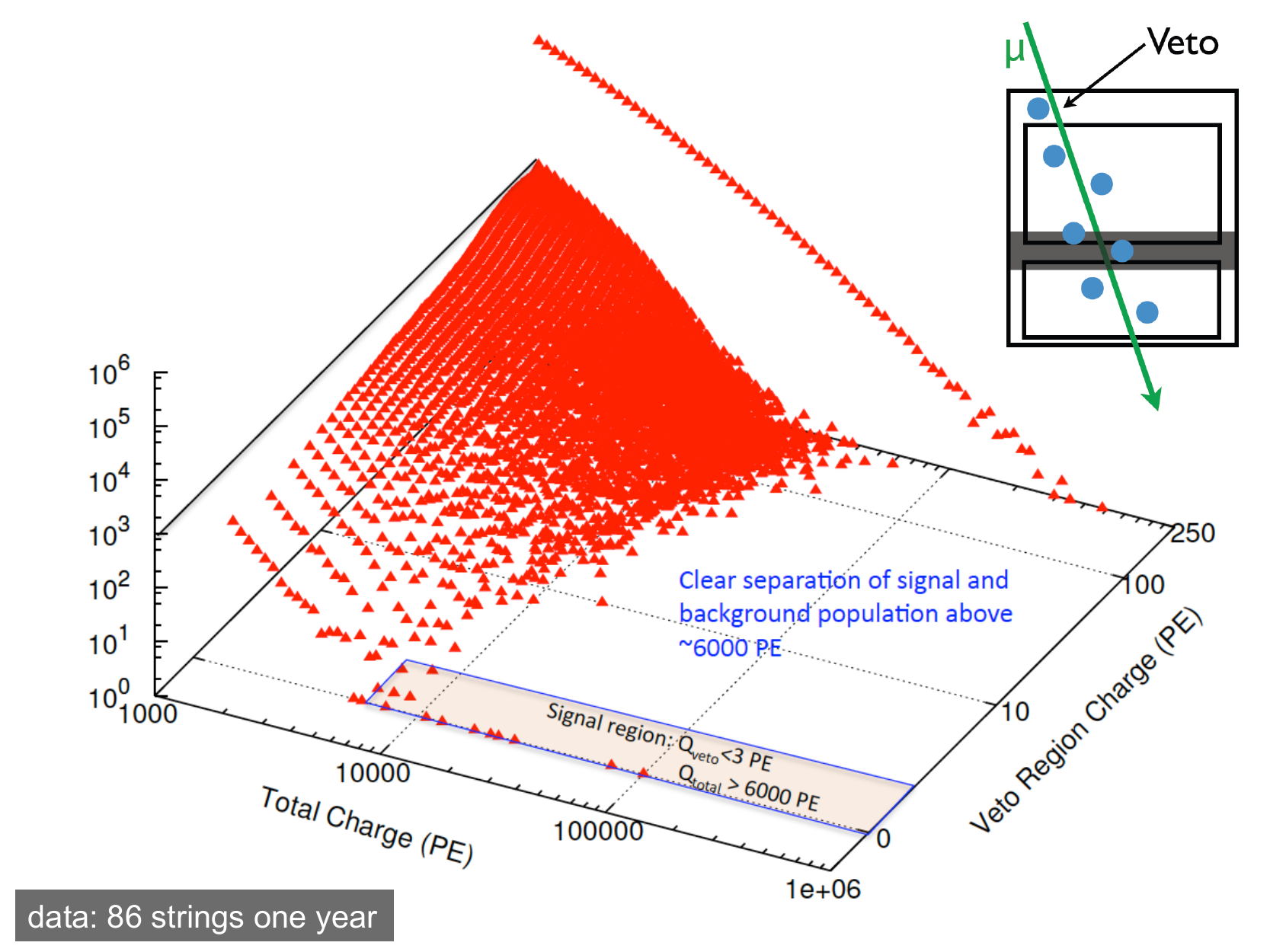}
\caption{One year of IceCube data from its final 86-string configuration showing number of events as a function of the total number of photoelectrons and the number present in the veto region. The signal region requires  more than 6000 photoelectrons with less than three of the first 250 in the veto region of the detector. The signal, including nine events with reconstructed neutrino energy in excess of 100\,TeV, is clearly separated from the background.}
\label{hesedata}
%%\vspace{-15pt}
\end{center}
\end{figure}

Amazingly, the emergence of a component of cosmic neutrinos from a one-year sample of atmospheric muons and neutrinos triggered by the detector can be demonstrated by just plotting the data, as shown in Fig.~\ref{hesedata}.

The second (and in fact third) year of data yields statistically consistent samples of neutrinos with energies inconsistent with atmospheric origin and, additionally, showing no trace of an accompanying atomspheric shower in which they would have originated. 

Of the 28 events, 21 are showers with no evidence of a muon track and with energies measured to better than 15\%, although their directions are determined to 10-15 degrees only.  
The remaining seven events are muon tracks, which do allow for sub-degree angular reconstruction; however, only a lower limit on their energy can be established because of the unknown fraction carried away by the exiting muon. Furthermore, the lower energy muon-like events from above include four that start near the detector boundary and are
consistent with the expected background of atmospheric muons.  The expected background
is $10.6^{+5.0}_{-3.6}$ events, which has comparable contributions from atmospheric muons and atmospheric neutrinos.
The sample of 28 events represents an excess of more than 4 standard deviations
above background. 

\begin{figure}[htpb]
\begin{center}
%%\vspace{-25pt}
\includegraphics[width=\linewidth]{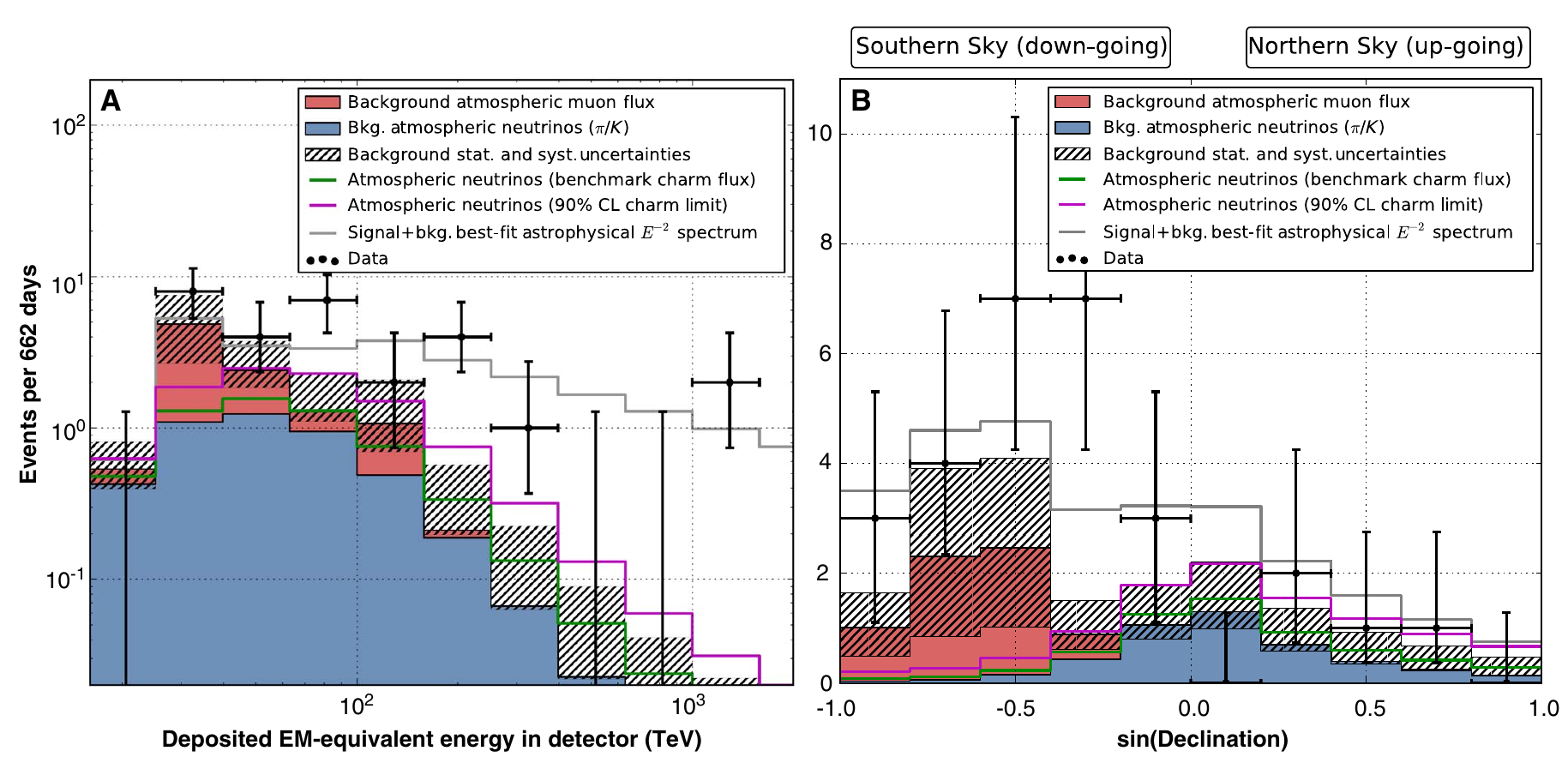}
\caption{Distribution of the deposited energies (left) and declination angles (right) of the observed events compared to model predictions. Energies plotted are in-detector visible energies, which are lower limits on the neutrino energy. Note that deposited energy spectra are always harder than the spectrum of the neutrinos that produced them due to the neutrino cross section increasing with energy. The expected rate of atmospheric neutrinos is based on Northern Hemisphere muon neutrino observations. The estimated distribution of the background from atmospheric muons is shown in red. Due to lack of statistics from data far above our cut threshold, the shape of the distributions from muons in this figure has been determined using Monte Carlo simulations with total rate normalized to the estimate obtained from our in-data control sample. Combined statistical and systematic uncertainties on the sum of backgrounds are indicated with a hatched area. The gray line shows the best-fit $E^{-2}$ astrophysical spectrum with all-flavor normalization (1:1:1) of $E_\nu^2 \frac{dN}{dE_\nu}=3.6\times10^{-11}\,\rm TeV\,cm^{-2}\,s^{-1}\,sr^{-1}$ and a spectral cutoff of 2 PeV.}
\label{hese_energy_zenith}
\end{center}
\end{figure}

The energy and zenith angle dependence of the 28 events are shown in Fig.~\ref{hese_energy_zenith}.  There is a significant excess of events above
$100$~TeV compared to the background expectation.  Both the energy and zenith angle dependence
observed is consistent with what is expected for a flux of neutrinos produced by cosmic accelerators.  The flavor composition of the flux is, after corrections for the acceptances of the detector to the different flavors, consistent with 1:1:1 as anticipated for a flux originating in cosmic sources.

The large errors on the background are associated with the possible presence of a neutrino component originating from the production and prompt leptonic decays of charmed particles in the atmosphere. Such a flux has not been observed so far. While its energy and zenith angle dependence are known, its normalization is not; see Fig.~\ref{discovery} for one attempt at calculating the flux of charm origin. Neither the energy, nor the zenith angle dependence of the 28 events observed can be described by a charm flux, and, in any case, fewer than 3.4 events are allowed at the 1\,$\sigma$ level by the present upper limit on a charm component of the atmospheric flux set by IceCube itself.  

Fitting the data to a superposition of extraterrestrial neutrinos on an atmospheric background yields a cosmic neutrino flux of
\begin{equation}
E_\nu^2 \frac{dN}{dE_\nu}=3.6\times10^{-8}\,\rm GeV\,cm^{-2}\,s^{-1}\,sr^{-1}
\label{eq:heseflux}
\end{equation}
for the sum of the three neutrino flavors. This is the level of flux anticipated for neutrinos accompanying the observed cosmic rays.

\begin{figure}[htpb]
\begin{center}
%%\vspace{-25pt}
\includegraphics[width=0.9\linewidth]{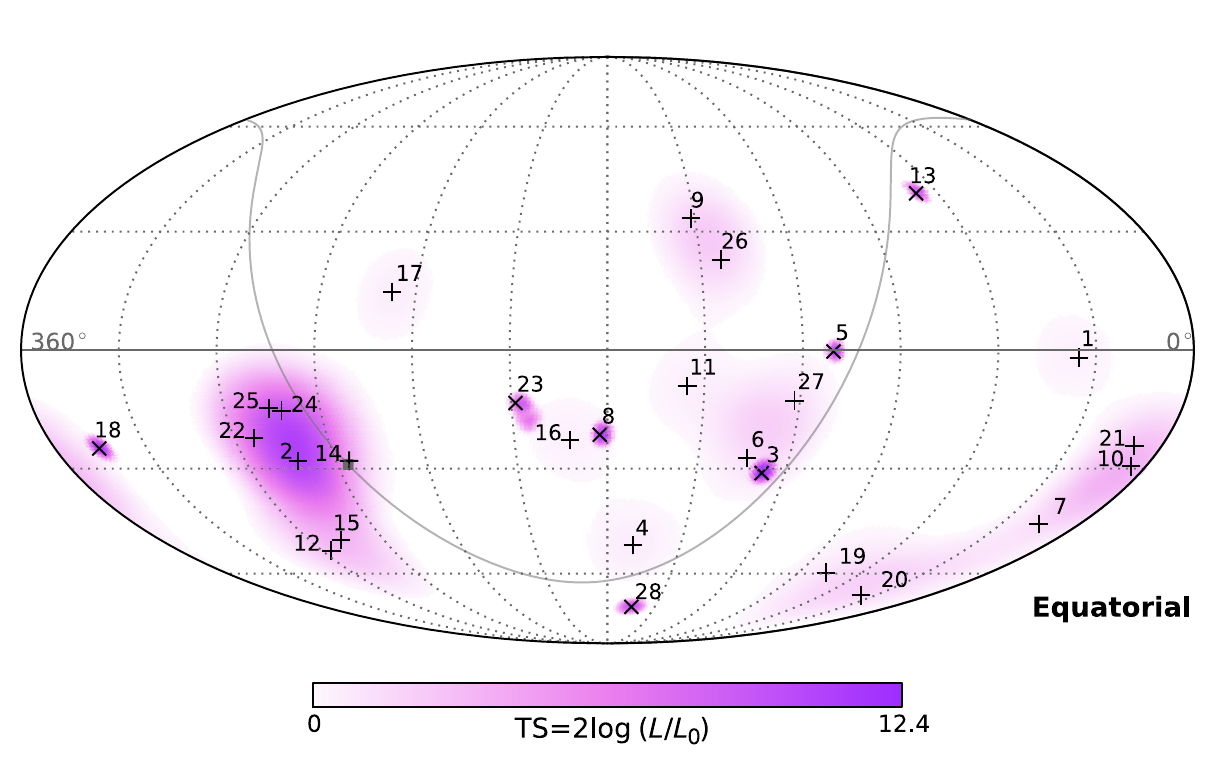}
\caption{Sky map in equatorial coordinates of the test statistic (TS) that measures the probability of clustering among the 28 events. The most significant cluster consists of five events---all showers and including the second-highest energy event in the sample---with a final significance of only 8\%. The Galactic plane is shown as a gray line with the Galactic center denoted as a filled gray square. Best-fit locations of individual events are indicated with vertical crosses (+) for showers and angled crosses (x) for muon tracks.}
\label{hesemap}
\end{center}
\end{figure}

So, where do the neutrinos come from? A map of their arrival directions is shown in Fig.~\ref{hesemap}. A test statistic $TS=2 \times log{L/L_0}$ was used, where L is the signal plus background maximized likelihood and $L_0$ is the background only likelihood obtained by scrambling the data. No significant spot on the sky was found when compared to  randomized pseudoexperiments. Repeating the analysis for showers only, a hot spot appears at RA=281 degrees and dec=23 degrees, close to the Galactic center. After correcting for trials, the probability corresponding to its TS is 8\%. Searches
were also made for clustering of the events in time and for a possible correlation with the times of observed GRBs. No statistically significant correlation was found. Fortunately, more data is already available, and the analysis, performed blind, can be optimized for searches of future data samples.

\section{Outlook}

This conference covering cosmology, astroparticle physics, and neutrino physics provided us with an overview of the leading experimental results and the considerable intellectual challenges that have emerged. The message is clear: there is some very basic physics that is not part of our present understanding of the Standard Model and, possibly, of gravity as well. 

In this talk, I have emphasized that the same message has been reinforced by the discovery of the Higgs. I would like to underscore two points raised in the introduction. The search for new physics is not the province of only accelerator or only non-accelerator experiments, we should pursue both. This is especially true because we have very little guidance from theory at this point; supersymmetry at the natural scale has not shown up so far. And there is ample evidence in the past of great synergy: the discovery of helium in the sun, the formation of carbon by three alpha particles in stars, and the discovery of neutrino mass with natural neutrino beams. 

Clearly, the LHC has the potential to detect dark matter, and particle astrophysics experiments are unlikely to compete for revealing signatures of new physics far beyond the TeV scale. However, astroparticle detectors observe neutrinos with energies and in numbers that are unmatched by accelerators on Earth. Focused on the search for the cosmic ray accelerators, we often overlook their potential for particle physics discoveries. One can dream of collecting large samples of neutrinos with energies in excess of a million TeV using radio telescopes embedded in polar ice. The measurement of their cross sections would probe electroweak theory to scales unlikely to ever be matched by earthly accelerators. The beam exists: the cosmic rays we observe must interact with microwave photons to produce pions that decay into neutrinos of such energies.

The road to future success is to keep up the impressive build-up of instrumentation that we have witnessed over the last decade on all fronts and not get depressed over whether the LHC should, or should not, have discovered anything new by this time; it already discovered the Higgs and has some way to go in energy and luminosity.

\Acknowledgements
I would like the thank the speakers for inspiring talks, not forgetting the barrage of ideas that we witnessed the day of the parallel session. This was a very enjoyable meeting as well. For their inspiration and organizational success, I thank my friends in Hawaii, especially Jason Kumar and Xerxes Tata.  This research was supported in part by the U.S. National Science Foundation under Grants No.~OPP-0236449 and PHY-0969061 and by the University of Wisconsin Research Committee with funds granted by the Wisconsin Alumni Research Foundation.

\end{document}